\documentclass{PoS}

\title{Playing with the kinetic term in the HMC}

\ShortTitle{Playing with the kinetic term in the HMC}

\usepackage{booktabs}
\usepackage{graphicx}
\usepackage{caption}
\usepackage{subcaption}
\usepackage{multirow}

\author{\speaker{Alberto Ramos}\\
        NIC, DESY. Platanenallee 6, 15738 Zeuthen, Germany\\
        E-mail: \email{alberto.ramos@desy.de}}

\abstract{The HMC algorithm, combining the advantages of molecular
  dynamics and 
Monte-Carlo methods, is the most efficient algorithm to simulate QCD
including the effects of sea quarks. In the standard approach momentum
fields are generated with a Gaussian probability density. In this work
I will explore another possibility. By using a Lorentz distribution one
can dynamically impose a cutoff in the rate of change of the
coordinates that potentially could have a better behavior. 
I will present some results in pure $SU(2)$ gauge theory.
\vspace{2cm}
\begin{flushright}
DESY 12-236
\end{flushright}
}

\FullConference{The 30th International Symposium on Lattice Field Theory\\
		 June 24 - 29,  2012\\
		 Cairns, Australia}

\begin{document}

\section{Introduction}

In a QCD lattice simulation the fermionic degrees of freedom are
integrated out, producing a non local effective gauge
action that increases dramatically the cost of evaluating any
update. Algorithms based on local updates are terribly inefficient 
in simulating these kinds of systems. The
Hybrid Monte Carlo (HMC) algorithm~\cite{Duane:1987de} combines 
a molecular dynamics update scheme with a Metropolis-Hastings 
accept-reject step to produce global update proposals but keeping a
high acceptance probability. These unique characteristics make the
HMC algorithm the only choice for large-scale QCD simulations. 

The global update proposals of the HMC come from a Hamiltonian flow
\begin{equation}
  H(p,q) = \frac{p^2}{2} + S[q],
\end{equation}
where we represent by $q$ the fields of the system that we want to
study and by $p$ the additional conjugate momenta variables. These
additional conjugate momenta follow a gaussian probability
distribution. In physical systems this kinetic term is a
consequence of Galilean and rotational 
invariance. But these are symmetries that we do not have to obey in
our lattice simulations. We can choose different kinetic terms if they
produce a more effective sampling.  

In this work we are going to analyze two different ways to explore
phase space by changing the kinetic term. The standard kinetic term
produces via the equation of motion $\dot q = \frac{\partial
  H}{\partial p} = p$ a rate of change of the variables that also
follows a Gaussian 
distribution. In our update process the change in the values of the
coordinates is therefore centered at zero with a few variables
changing by a large amount. As an alternative we are going to use the
Hamiltonian 
\begin{equation}
  H(p,q) = \log\left(1+\frac{p^2}{2\gamma^2}\right) + S[q]
\end{equation}
making our momenta follow a Lorentz distribution $\propto \frac{1}{1
  +{p^2}/{2\gamma^2}}$. These
conjugate momenta are also centered around zero, but their
fluctuations don't have a typical scale. This may seem a bad idea because
very large momenta will kick the variables away, thus making our
numerical integration of the equations of motion difficult. But the
situation is just the opposite. The equation of motion for our variables
reads now
\begin{equation}
  \dot q = \frac{\partial H}{\partial p} = \frac{2p}{\gamma^2+p^2}
\end{equation}
and therefore this kinetic term effectively imposes a dynamic cutoff
in the rate of change of our variables: $\dot q$ can not be bigger
than $1/\gamma$. Very large
momenta or forces are not dangerous in the modified algorithm because
they simply ``freeze'' the conjugate link variables instead of
kicking them. Moreover, the distribution of
$\dot q$ peaks close to its maximum possible value. In this way we
produce a motion in phase space that is on average faster by avoiding
very large updates. 

We are going to compare the standard HMC algorithm and the modified one
with a Lorentz kinetic term (we will call it HMCL) in a pure 
$SU(2)$ gauge theory. 

\section{Implementation in pure $SU(2)$ gauge theory}

We are going to consider pure $SU(2)$ gauge theory (the extension to
other gauge groups is straightforward). The links
$U_\mu(x)\in SU(2)$ take values in the group, and elements of the
algebra $\pi\in\mathfrak{su}(2)$ can be written as 
\begin{equation}
  \pi = \sum_{a=1}^3 \pi^a T^a
\end{equation}
where $T^a$ are the $SU(2)$ generators (the $T^a=\sigma^a/2$ with
$\sigma^a$ being the Pauli matrices).

We are going to study the plaquette action
\begin{equation}
  S[U] = \frac{\beta}{2}\sum_{p}Tr(1-U_p),
\end{equation}
where $U_p$ is the product of the links in a simple plaquette and the
sum is over all oriented plaquettes. For each link $U_\mu(x) \in SU(2)$ we
introduce an algebra-valued conjugate momenta $\pi_\mu(x) \in
\mathfrak{su}(2)$. The Hamiltonian that defines the standard HMC
algorithms is given by
\begin{equation}
  H(\pi,U) = \sum_{x,\mu,a}\frac{\left[\pi^a_\mu(x)\right]^2}{2} +
  \frac{\beta}{2}\sum_{p}Tr(1-U_p)
\end{equation}
with equations of motion
\begin{eqnarray}
  \label{eq:f1}
  \dot \pi_\mu^a(x) &=&  -\frac{\partial H(\pi,U)}{\partial
    U_\mu(x)} = 
  -\frac{\beta}{4}Tr\left[T^a
    {U_\mu(x)}\sum_{\nu\neq\mu}{
      M_{\mu\nu}(x)} \right] \\ 
  \dot U_\mu(x) &=&  \frac{\partial H(\pi,U)}{\partial
    \pi_\mu(x)} = \pi_\mu(x) U_\mu(x) 
\end{eqnarray}
where $M_{\mu\nu}(x)$ are the sum of ``staples'' of Wilson lines from
$x$ to $x+\mu$. The Hamiltonian of the modified (HMCL) algorithm
is given by\footnote{Note that the sum in the internal ``color''
index $a$ is \emph{outside} the $\log$.}
\begin{equation}
  \label{eq:hl}
    H_L(\pi, U) = \sum_{x,\mu, a}\log\left[1+\left({\pi_\mu^a(x)}/{\gamma}\right)^2\right]
    + \frac{\beta}{2}\sum_p Tr(1-U_p).
\end{equation}
Only the equation of motion for $U_\mu(x)$ is modified, that now reads
\begin{equation}
  \dot U_\mu(x) =  \frac{\partial H(\pi,U)}{\partial
    \pi_\mu(x)} =  \sum_a\left[\frac{2\pi_\mu^a(x)}{\gamma^2+
            (\pi_\mu^a(x))^2}\, T^a\right] U_\mu(x) 
\end{equation}
while the equation of motion for the momenta (the computation of the 
forces) remains unchanged. 

Together with the change in the equation of motion of $U_\mu(x)$ we
need two additional changes to our standard HMC algorithm in order to
achieve the desired equilibrium properties. First the momentum refresh
has to be done using a Lorentz distribution\footnote{Generating random
  samples of this distribution is straightforward, since this
  probability density function is easily integrable.} with scale
$\gamma$. Second, one has to accept the update with a probability
$\min(1,e^{-\Delta H_L})$ where $H_L$ is given by
  equation~(\ref{eq:hl}). 

\subsection{Details of the simulations and observables}

We simulate at three different values of the coupling $\beta = 2.4,
2.5, 2.6$ that should span roughly a factor two change in the
scale ($a\sim 0.06-0.12$ fm), and two volumes $V = 12^4, 16^4$. The
molecular dynamics equations of motion  are integrated using the
Omelyan integrator~\cite{Omelyan:2003aa} with the standard value of 
$\lambda=0.185$. 


The {HMCL} algorithm has an extra parameter $\gamma$ to
tune. This parameter can be interpreted as a rescaling in the
molecular dynamics time and in this sense is similar to a mass term in
the 
normal {HMC} algorithm. Obviously the smaller this mass
parameter is, the faster one will move trough phase space, but this has
a cost in terms of acceptance ratio. Here we will tune the $\gamma$
parameter to achieve similar acceptance ratios on both the
{HMC} and the {HMCL} setups. This is achieved with
$\gamma = 0.8$.

We will measure Wilson loops of sizes $1\times 1$ (the plaquette), and
also $2\times 2$ and $4\times 4$. We will also look at the topological
charge. We use the clover definition of $F_{\mu\nu}$ to compute 
\begin{equation}
  Q = \frac{1}{32\pi^2}\sum_x
  \varepsilon_{\mu\nu\rho\sigma}F_{\mu\nu}(x)F_{\rho\sigma}(x).
\end{equation}
It is well known that Monte-Carlo configurations contain large UV
fluctuations that hide large-scale structures like
the topological charge. We will apply a smoothing process to our gauge
configurations, in particular an under relaxed cool~\cite{Michael:1994rt}
\begin{equation}
  U^{(n+1)}_\mu(x) \longrightarrow  \mathcal P\{ \alpha U^{(n)}_\mu(x) +
  \sum_\nu 
  M_{\mu\nu}(x) \}
\end{equation}
in which each link is averaged with its six ``staples'' and then
projected back to the group. We apply 60 steps with 
$\alpha=2$. 

\section{Results}

Both the HMC and the HMCL algorithm allow us to measure observables
in a sample of the equilibrium distribution $\left\{ O_\alpha(t);
  t=1,\dots,N\right\}$ (the index $\alpha = 1,\dots, N_{\rm obs}$
labels the observables). These measurements are statistically
correlated, and the autocorrelation function quantifies these
correlations 
\begin{equation}
  \Gamma_{\alpha}(t) =\frac{1}{N}\sum_{t=1}^N \left(O_\alpha(t) - \overline
    O_\alpha\right)\left(O_\alpha(0) - \overline O_\alpha\right)
  \rangle,  
\end{equation}
where $\overline O_\alpha$ is the statistical mean of the observable
$\alpha$. 

These correlations have to be taken into account when computing the
statistical error of an observable (see for
example~\cite{Wolff:2003sm}). In our case it is given by
\begin{equation}
  \label{eq:err}
  \sigma_\alpha = \sqrt{\frac{2v_\alpha\tau_{{\rm int},\alpha}}{N}}
\end{equation}
where $v_\alpha$ is the variance of the observable $\alpha$ (that is,
$\Gamma_\alpha(0)$), and $\tau_{{\rm int},\alpha}$ is the integrated
autocorrelation time of the observable, defined as
\begin{equation}
  \tau_{{\rm int},\alpha} = \frac{1}{2} +
  \frac{1}{v_\alpha}\sum_{t=1}^\infty  
  \Gamma_\alpha(t).
\end{equation}

\begin{figure}
  \centering
  \begin{subfigure}[b]{0.45\textwidth}
    \includegraphics[width=\textwidth]{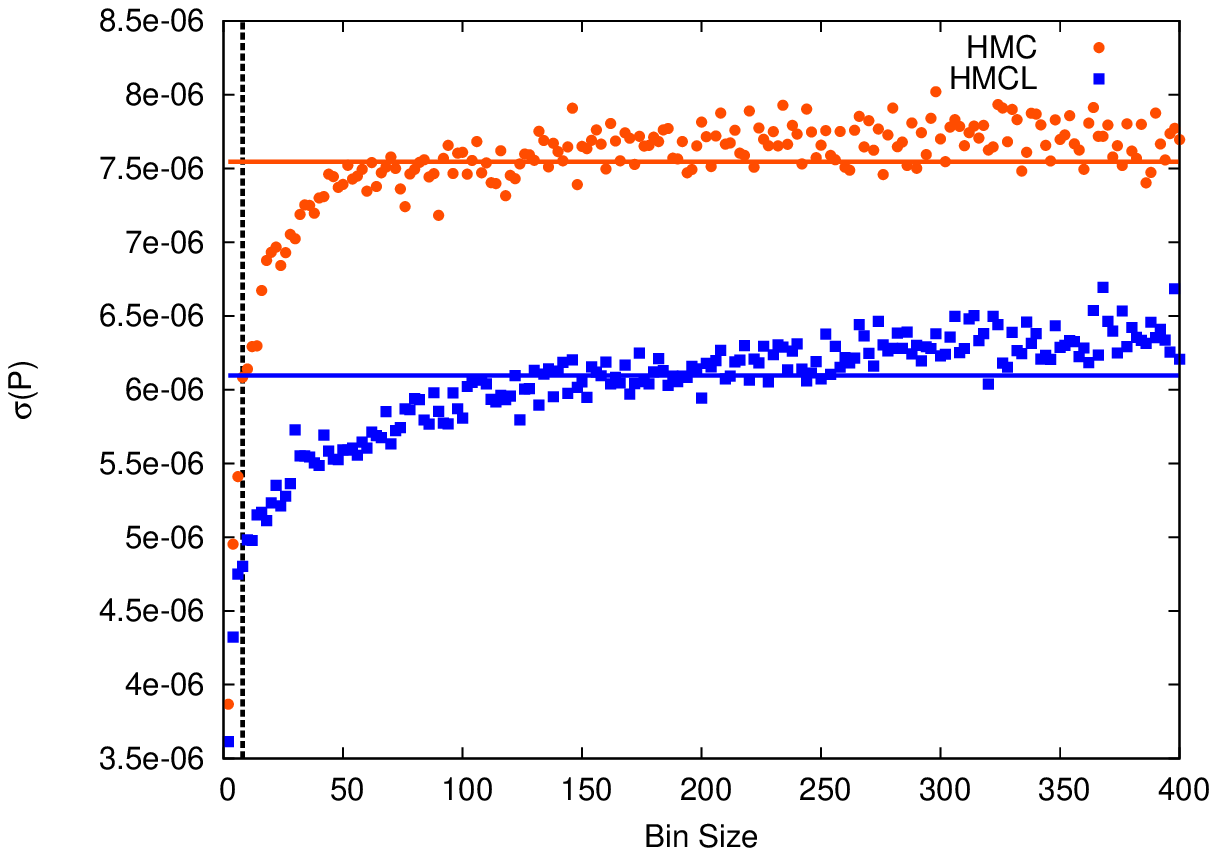}
    \caption{Plaquette}
  \end{subfigure}
  \begin{subfigure}[b]{0.45\textwidth}
    \includegraphics[width=\textwidth]{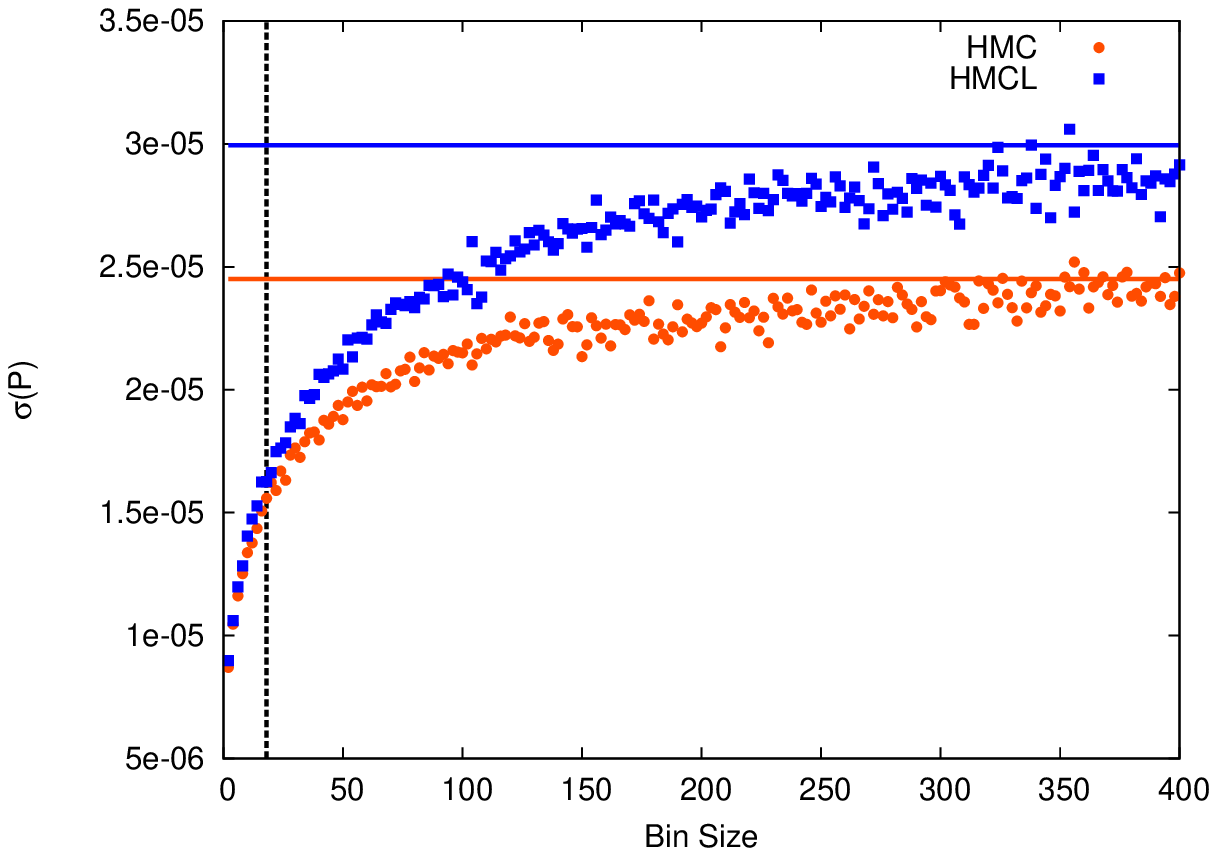}
    \caption{$\langle W_{4\times 4}\rangle$}
  \end{subfigure}
  \caption{Error versus the bin size. The vertical dashed line
    represents $2\tau_{\rm int}$ and is plotted for reference
    only. The horizontal lines corresponds to the error computed by
    computing $\tau_{\rm int}$ (see text for more details).}
  \label{fig:bin}
\end{figure}

One can proceed through the previous formulas to compute the
uncertainties of the observables, or one can use the popular
alternative of binning methods, in which data is averaged over
sections of the Monte Carlo history of large enough length so that
these measurements can be considered independent. In our particular
case, with very long Monte Carlo histories both methods are
straightforward to apply and give the same results (see
figure~\ref{fig:bin}), but when this is not the case the discussion is
more subtle~\cite{Schaefer:2010hu}. 

Since in our setup both algorithms need the same computing
time to produce an additional member of the Markov chain, and the
acceptance probabilities in both algorithms are similar, the question
of which algorithm performs better is answered by looking at the
algorithm that ``decorrelates'' faster, producing smaller
uncertainties. Table~\ref{tab:res} summarizes the results for
different observables. 
\begin{table}
  \centering
  {\footnotesize
  

\begin{tabular}{llll|lll}
  \bottomrule
  \multicolumn{4}{c}{$V=12^4$} & 
  \multicolumn{3}{|c}{$V=16^4$} \\
   $\langle \dots\rangle$ & $\beta=2.4$ & $\beta=2.5$ & $\beta=2.6$ & 
    $\beta=2.4$ & $\beta=2.5$ & $\beta=2.6$  \\ 
  \hline
  \multirow{2}{*}{$W_{1\times 1}$} &  1.259998(11)
  & 1.3040202(99) & 1.3401848(77)  
  & 1.259995(16)  & 1.303919(13)  & 1.340050(12) \\
  & 1.260004(11)  & 1.3040278(93) & 1.3401944(62) 
  & 1.259982(15)  & 1.303952(11)  & 1.3400554(94) \\
  \hline
  \multirow{2}{*}{$W_{2\times 2}$} & 0.444850(25) &
  0.516329(25) &   0.575913(19)  
  & 0.444825(36) & 0.515973(31) & 0.575331(32)  \\
  & 0.444865(30) & 0.516315(30) & 0.575928(20) 
  & 0.444790(40) & 0.516035(32) & 0.575352(38) \\
  \hline
  \multirow{2}{*}{$W_{4\times 4}$} & 0.026182(12) & 0.048300(25)
  & 0.072296(21)  
  & 0.026167(17)  & 0.048031(25) & 0.071383(41) \\ 
  & 0.026194(16)  & 0.048362(29) & 0.072288(26)
  & 0.026157(20)  & 0.048061(30) & 0.071385(57)  \\
  \hline
  \multirow{2}{*}{$Q^2$} & 2.650(12) & 0.8523(87) & 0.1232(40) 
  & 8.451(83) & 3.055(56) & 0.808(36)  \\
  & 2.661(16) & 0.849(12) & 0.1266(51)
  & 8.48(11)  & 3.068(78) & 0.786(46) \\
  \toprule
\end{tabular}

}
  \caption{Observables measured in the run with their
    uncertainties. The first line are results of the standard HMC
    algorithm, whereas the second line corresponds to the results of
    the HMCL algorithm. As
    one can see, the agreement between both algorithms is almost
    perfect.  } 
  \label{tab:res}
\end{table}

One observe that the results are very similar between the two
algorithms, with a difference in the uncertainties in the range $\pm
30\%$, and this difference seem to be constant for all values of
$\beta$ and the volume (i.e. neither algorithm seems to have a better
\emph{scaling} than the other). Nevertheless  all observables but the
plaquette decorrelate faster with the standard kinetic term
of the HMC algorithm than with the modified version. 

The biggest difference is found in the topological susceptibility. It
is known that simulations suffer from a critical slowing down when
approaching the 
continuum limit, and that this problem is more severe for topological
quantities~\cite{Schaefer:2010hu} and recently the solution of simulating
with open boundary conditions have been
proposed~\cite{Luscher:2011kk}. We wanted to see if the previous
conclusions remain unchanged when simulating with open boundary
conditions. 

\subsection{Simulations with open boundary conditions}

\begin{figure}
  \centering
  \begin{subfigure}[b]{0.45\textwidth}
    \includegraphics[width=\textwidth]{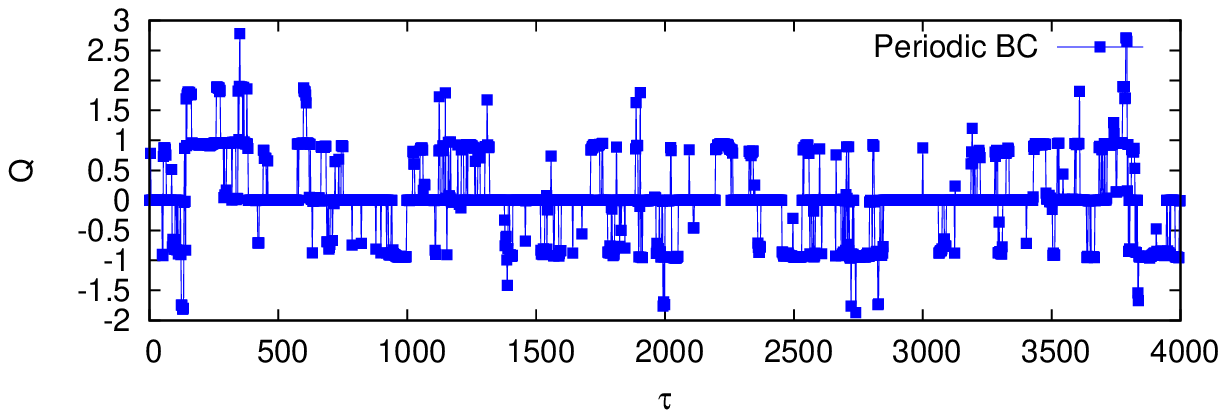}
  \end{subfigure}
  \begin{subfigure}[b]{0.45\textwidth}
    \includegraphics[width=\textwidth]{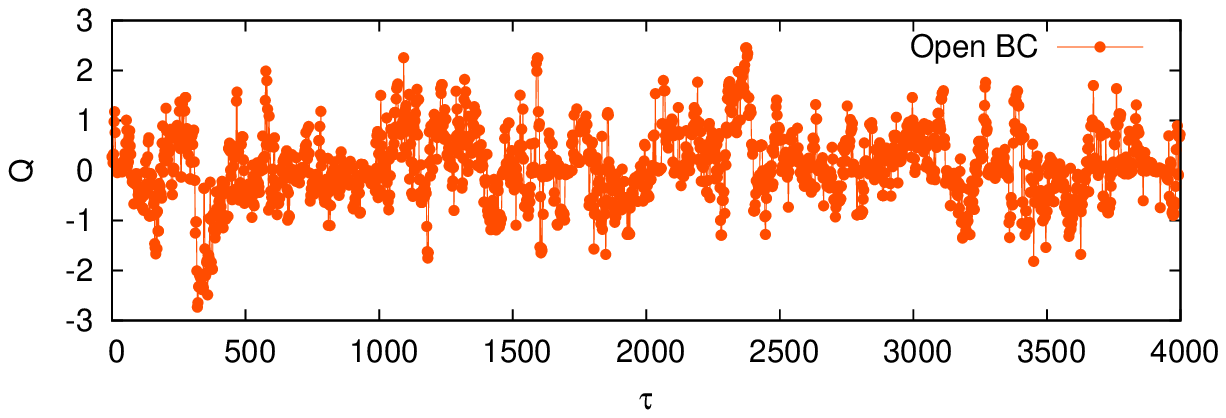}
  \end{subfigure}
  \caption{Sample of the Monte Carlo history of the topological charge
with open and periodic boundary conditions for the volume $V=12^4$ and
$\beta = 2.5$.}
\label{fig:hist}
\end{figure}

By using open (Neumann) boundary conditions in the time direction the
topological charge can go in and out the lattice. This has the
effect observed in the histories of Fig.~\ref{fig:hist}, where by
simple observations one can see that the measurements of the
topological charge are less correlated in the runs with open boundary
conditions. 

Nevertheless, as the results of the Table~\ref{tab:obc} show, the
conclusions remain unchanged for the runs performed with open boundary
conditions: There is not much difference between both algorithms, and
only the plaquette decorrelates faster with the HMCL algorithm. 

\begin{table}
  \centering
  {\footnotesize
  

\begin{tabular}{llll|lll}
  \bottomrule
  \multicolumn{4}{c}{$V=12^4$} & 
  \multicolumn{3}{|c}{$V=16^4$} \\
   $\langle \dots\rangle$ & $\beta=2.4$ & $\beta=2.5$ & $\beta=2.6$ & 
    $\beta=2.4$ & $\beta=2.5$ & $\beta=2.6$  \\ 
  \hline
  \multirow{2}{*}{$W_{1\times 1}$} &  1.190795(20)
  & 1.235107(17)  & 1.273033(16)
  & 1.207584(26)  & 1.251767(20) & 1.28924(16) \\
  & 1.190795(19)  & 1.235084(17) & 1.27298(13)
  & 1.207604(22)  & 1.251780(19) & 1.28924(15)   \\
  \hline
  \multirow{2}{*}{$W_{2\times 2}$} & 0.363959(38) &
  0.425108(35) &   0.478480(30)
  & 0.383586(53)  & 0.447091(42)  & 0.501832(33)  \\
  & 0.363959(44) &  0.425084(41)  & 0.478392(34)
  & 0.383622(57) &  0.447093(49)  & 0.501830(41) \\
  \hline
  \multirow{2}{*}{$W_{4\times 4}$} & 0.015822(16) & 0.029544(23)
  & 0.045165(25)
  & 0.018301(22)  & 0.034012(31) & 0.051318(29) \\ 
  & 0.015826(19)  & 0.029555(27) & 0.045103(34)
  & 0.018362(26)  & 0.033976(37) & 0.051365(39)  \\
  \hline
  \multirow{2}{*}{$Q^2$} &  1.334(11) & 0.4849(71) & 0.1409(39)
  & 5.398(96) & 1.840(41) & 0.599(27)  \\
  & 1.336(14) & 0.4850(90)  & 0.1453(68)
  & 5.25(10)  & 1.933(57) & 0.598(34) \\
  \toprule
\end{tabular}

}
  \caption{Observables measured in the runs with open boundary
    conditions. The first line are results of the standard HMC
    algorithm, whereas the second line corresponds to the results of
    the HMCL algorithm. These results draw the same conclusions as the
    runs with periodic boundary conditions. } 
  \label{tab:obc}
\end{table}

\section{Conclusions}

By changing the kinetic term in the Hybrid Monte Carlo we have
investigated another way to explore the phase space. The equations of
motion of the modified Hamiltonian impose dynamically a cutoff in the
rate of change of the coordinates, avoiding the effects of either
large forces or momenta. In this way we pretend a faster exploration
of phase space. This modification does not need any
change in the evaluation of the molecular dynamics forces, and is thus
easy to implement over an existing HMC simulation code.

We have compared the standard and modified kinetic terms in pure
$SU(2)$ gauge theory by generating long Markov chains at two
different volumes, three values of the coupling and different boundary
conditions (periodic and open). By tuning both algorithms to use the same 
computer time per trajectory, and to have the same acceptance ratio, a
meaningful comparison between both algorithms can be done by simply
looking at which algorithm decorrelates faster the measurements, that
would translate in smaller uncertainties at the same computational
cost. We have investigated different observables: Wilson loops of
different sizes and the topological susceptibility. 

We find that both algorithms perform rather similarly, with a difference
in the uncertainties of around $\pm 30\%$ for all values of the
coupling or of the volume (that is: both algorithms have a similar
scaling). It seems that the modified 
kinetic term tends to decorrelate faster the plaquette (a very local
observable), while larger Wilson loops or the topological
susceptibility tends to decorrelate faster with the standard kinetic
term. 

Having very similar performances, and noting that usually one is
interested in quantities less local than the plaquette, there seems to
be no advantage, but a worsening in performance, in the use of the
modified kinetic term in terms of producing ``more independent''
samples of the equilibrium distribution. 

Still, it is not in the pure gauge theory where the effect of large
forces is problematic, rather in the simulation of dynamical fermions. 
These simulations can benefit from having a cutoff in the
rate of change of the coordinates. This point needs
further investigations. 

\section*{Acknowledgments}

I would like to express my special thanks to the many
people who helped me in this project. This would have been impossible
without the discussions, code, time, and above all, the patience of
Patrick Fritzsch, Laurent Lellouch, Marina Marinkovic, Alfonso Sastre,
Hubert Simma, Rainer Sommer and Francesco Virotta. Thanks, guys!

\bibliography{/home/alberto/latex/math,/home/alberto/latex/campos,/home/alberto/latex/fisica,/home/alberto/latex/computing}

\end{document}